# Gate-based quantum computing for protein design

Mohammad Hassan Khatami[1], Udson C. Mendes[2], Nathan Wiebe[3,4,5], Philip M. Kim[1,3,6,*]

[1]Terrence Donnelly Centre for Cellular & Biomolecular Research, University of Toronto, Toronto, ON, Canada

[2]CMC Microsystems, Sherbrooke, QC, Canada

[3]Department of Computer Science, University of Toronto, Toronto, ON, Canada

[4]Pacific Northwest National Laboratory, Richland WA USA

[5]Department of Physics University of Washington, Seattle WA USA

[6]Department of Molecular Genetics, University of Toronto, Toronto, ON, Canada

*Corresponding Author

Email: pi@kimlab.org (PMK)

**Abstract**

Protein design is a technique to engineer proteins by permuting amino acids in the sequence to obtain novel functionalities. However, exploring all possible combinations of amino acids is generally impossible due to the exponential growth of possibilities with the number of designable sites. The present work introduces circuits implementing a pure quantum approach, Grover's algorithm, to solve protein design problems. Our algorithms can adjust to implement any custom pair-wise energy tables and protein structure models. Moreover, the algorithm's oracle is designed to consist of only adder functions. Quantum computer simulators validate the practicality of our circuits, containing up to 234 qubits. However, a smaller circuit is implemented on real quantum devices. Our results show that using $\mathcal{O}(\sqrt{N})$ iterations, the circuits find the correct results among all $N$ possibilities, providing the expected quadratic speed up of Grover's algorithm over classical methods (i.e., $\mathcal{O}(N)$).

**Author summary**

Protein design aims to create novel proteins or enhance the functionality of existing proteins by tweaking their sequences through permuting amino acids. The number of possible configurations, $N$, grows exponentially as a function of the number of designable sites ($s$), i.e., $N=A^s$, where $A$ is the number of different amino acids ($A=20$ for canonical amino acids). The classical computation methods require $\mathcal{O}(N)$) queries to search and find the low-energy configurations among $N$ possible sequences. Searching among these possibilities becomes unattainable for large proteins, forcing the classical approaches to use sampling methods. Alternatively, quantum computing can promise quadratic speed-up in searching for answers in an unorganized list by employing Grover's algorithm. Our work shows the implementation of this algorithm at the circuit level to solve protein design problems. We first focus on lattice model-like systems and then improve them to more realistic models (change in the energy as a function of distances). Our algorithms can implement various custom pair-wise energy tables and any protein structure models. We have used quantum computer simulators to validate the practicality of our circuits which require up to 234 qubits. We have also implemented a simple version of our circuits on real quantum devices. Our results show that our circuits provide the expected quadratic speed-up of Grover's algorithm.

**1. Introduction**

Protein design is a procedure to construct proteins with certain configurations to achieve novel functionality. In this regard, amino acids are mutated in the protein's sequence to find sets of residues that provide the lowest energy of the protein in the expected configuration. Using computational approaches, one could consider having "*s*" designable sites in the sequence and "*A*" different amino acids that could fill these sites, where *A=20* for the canonical amino acids. This will provide $A^s$ possible sets of amino acids to find the answer. Thus, the number of possible sets of amino acid sequences grows exponentially by increasing the designable sites.

In computer science terminology, the protein design is categorized under non-deterministic polynomial-time (NP)-hard problems[1,2]. The main characteristics of these problems are that finding their answer using conventional computers is either impossible or requires a great deal of computational time. However, the validity of a proposed answer could be evaluated easily by conventional computers[3]. Even for the simple hydrophobic-polar (HP)[4] protein lattice models containing only two types of residues (hydrophobic and polar), the protein design is shown to be in the class of NP-hard problems[5]. Statistical methods such as Markov chain Monte Carlo (MCMC)[6,7] are currently being used to solve NP-hard problems, including protein design problems, on conventional computers[8–10]. In these methods, the algorithm uses sampling techniques and probability distributions to find the answers among all possible sets of amino acids. However, since the probabilistic methods do not explore all sets, it is possible to miss some of the answer states.

Unlike conventional approaches, quantum computation techniques are expected to enhance solving the NP class of problems in their exact forms[3]. In recent years, there have been attempts to use quantum computers to solve NP-hard problems in protein studies, mainly focused on protein folding[11–14]. In these studies, hybrid quantum-classical algorithms employing gate-based quantum devices, such as the Quantum Approximate Optimization Algorithm (QAOA)[15] and the Variational Quantum Eigensolver (VQE)[16], as well as quantum annealing approaches [17] are implemented. As an example of gate-based approaches, Fingerhuth *et al.* [14] used the QAOA method to study the protein folding on a 9-residue protein using the *HP* energy model and a 3D lattice structure. Similarly, a version of the VQE approach is employed by Robert *et al.*[13] to study the protein folding of a 10-residue protein, Angiotensin, and a 7-residue neuropeptide. In the case of implementing quantum annealers for protein folding, Perdomo-Ortiz *et al.* [11] studied the folding of a 6-residue peptide using a fixed energy table in a 2D lattice model, employing 81 D-Wave's "superconducting quantum bits". Similarly, Babej *et al.*[12] have worked on folding a ten residue Chignolin protein and an eight residue Trp-Cage peptide in 2D and 3D lattice models, respectively. This study used 2048 superconducting quantum bits of D-Wave's 2000Q quantum annealer device. Regardless of the approach, the studies employing quantum computation for protein folding are mainly limited to peptides with only a handful of residues and a very simplified or limited number of amino acid types, e.g., the *HP* model. In the case of protein design, Mulligan *et al.*[18] have prepared a hybrid quantum-classical solver in Rosetta software[19], called *QPacker*, to address the protein design problem on D-Wave's 2000Q quantum annealer device. Despite the attempts to use quantum computation, to the best of our knowledge, there are no records of studies in which a pure quantum computational method based on a pure gate-based approach is used to investigate protein design problems.

This work introduces a procedure to build gate-based quantum circuits, employing Grover's algorithm[20], to solve protein design problems. Grover's algorithm is a fundamental and famous quantum computation algorithm that offers a quadratic speedup in finding answers in an un-sorted list over classical methods[20–23]. In general, Grover's algorithm is composed of four main parts: initialization, Grover's oracle, Grover's diffuser and measurement (Fig. 1-A). The initialization, the diffuser and the measurement steps are almost similar for all systems in this study. However, the oracle step varies depending on the complexity of the system and is the only step that requires auxiliary work qubits, in addition to the *n* qubits (Fig. 1-B).

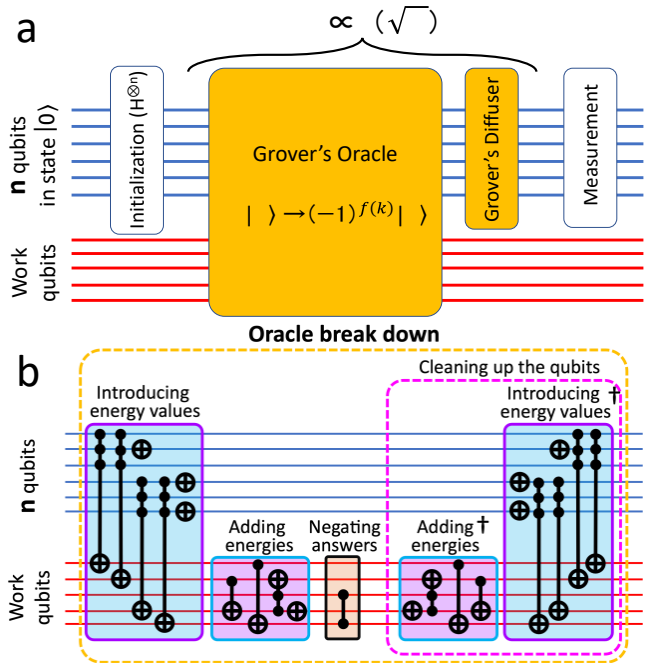

Figure 1: Schematic representation of our circuits.
A) Different steps of Grover's algorithm; B) Different sub-steps implemented in our Oracle. The parallel lines represent qubits. The blue lines show the *n* qubits and the red lines show the work qubits. In the oracle, if $|k\rangle$ is an answer state, $f(k) = 1$, otherwise it is 0. The MCX gates in the

In the initialization step, *n* qubits are allocated to create a superposition of $N = 2^n$ quantum states by applying Hadamard (H) gates i.e., $H^{\otimes n}$, representing all possible answer states in the system. This step is similar to other quantum algorithms such as Shor's algorithm[24], and the Deutsch-Jozsa algorithm[25].

In the oracle, the circuit is programmed to implement energies and do all necessary calculations to find the answer states (Fig. 1-B). This part of the algorithm inputs the general features of the structure (Fig. 2), pre-computed pair-wise interaction energy tables (Fig. 3), and a threshold energy ($E_{th}$) value (all discussed in detail later in the paper). In addition, pre-computed distance features are provided as input to the oracle depending on the system's complexity. The oracle is designed to use only the summation operation (and multiplication, technically a form of summation). First, the energy values are summed for each pair of interacting designable sites to find the total energy of the sequence. Then, it is subtracted from the $E_{th}$. If the result is negative, the oracle marks that sequence as an answer state by negating its amplitude (Fig. 1-B). Note that since the probability of each state is the only quantity of a quantum state that could be measured in the classic realm, the negative amplitude of the answer states in the oracle is not advantageous ($probability = (amplitude).(amplitude)^*$).

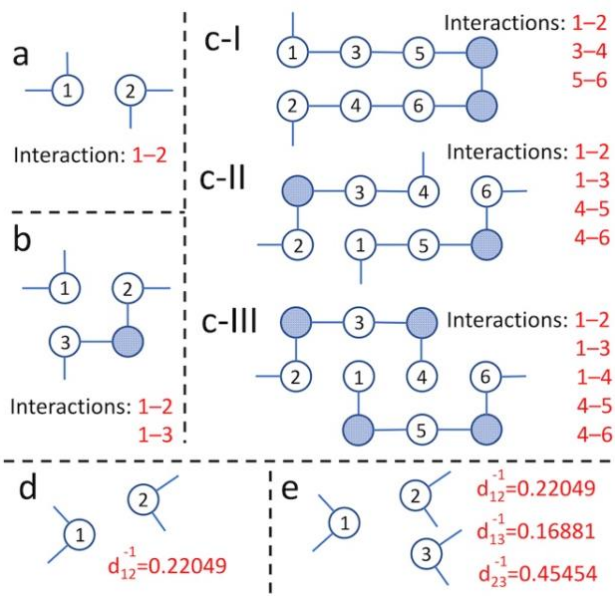

Figure 2: Schematic representations of protein models.
SP model with: a) Two designable sites; b) Three designable sites; c) Six designable sites with I) three, II) four, and III) five pair-wise interactions. MR model with: d) Two designable sites; e) Three designable sites. The designable sites are shown as circles with numbers. In a–c) the pattern of interactions between the sites are provided for each structure. The checkered circles are non-interacting residues. *In this study, there are no geometrical differences between residues, and all are being treated as identical beads represented with circles.* In d) and e) the $d_{ij}^{-1}$ is the corresponding distance reciprocal between designable sites *i* and *j*.

## a

| | H1 | H2 | Pol1 | Pol2 | Pos | Neg | X1 | X2 |
|---|---|---|---|---|---|---|---|---|
| H1 | -3 | -1 | +1 | 0 | +2 | +2 | 0 | 0 |
| H2 | -1 | -2 | 0 | +1 | +1 | +1 | 0 | 0 |
| Pol1 | +1 | 0 | -3 | -2 | -1 | -1 | 0 | 0 |
| Pol2 | 0 | +1 | -2 | -3 | +1 | 0 | 0 | 0 |
| Pos | +2 | +1 | -1 | +1 | +4 | -4 | 0 | 0 |
| Neg | +2 | +1 | -1 | 0 | -4 | +3 | 0 | 0 |
| X1 | 0 | 0 | 0 | 0 | 0 | 0 | -1 | 0 |
| X2 | 0 | 0 | 0 | 0 | 0 | 0 | 0 | 0 |

## b

| | H1 | H2 | Pol1 | Pol2 | Pos | Neg | X1 | X2 |
|---|---|---|---|---|---|---|---|---|
| H1 | -3.2 | -1.3 | +1.1 | +0.59 | +2.45 | +2.5 | 0 | 0 |
| H2 | -1.3 | -2.2 | 0 | +1 | +1.325 | +1.5 | 0 | 0 |
| Pol1 | +1.1 | 0 | -3.7 | -2 | -1 | -1 | 0 | 0 |
| Pol2 | +0.59 | +1 | -2 | -3.51 | +1.5 | 0 | 0 | 0 |
| Pos | +2.45 | +1.325 | -1 | +1.5 | +4.2 | -4.3 | 0.5 | 0 |
| Neg | +2.5 | +1.5 | -1 | 0 | -4.3 | +3.3 | 0 | 0 |
| X1 | 0 | 0 | 0 | 0 | 0.5 | 0 | -1.5 | 0 |
| X2 | 0 | 0 | 0 | 0 | 0 | 0 | 0 | 0.1 |

## c

| Res. | Binary ID | Res. | Binary ID |
|---|---|---|---|
| H1 | → 000 | Pos | → 100 |
| H2 | → 001 | Neg | → 101 |
| Pol1 | → 010 | X1 | → 110 |
| Pol2 | → 011 | X2 | → 111 |

## d

| | H | P |
|---|---|---|
| H | -1 | 0 |
| P | 0 | 0 |

| Res. | Binary ID |
|---|---|
| H | → 0 |
| P | → 1 |

Figure 3: Energy tables to represent the pair-wise interactions in our systems and the binary representation of residues. Energy tables for: a) The SP model; b) The MR model. c) Binary representation of residues in the energy table. d) Energy table and the residue representations for the HP protein model. In a–c) *H1* and *H2* represent two types of hydrophobic residues, *Pol1* and *Pol2* represent two types of polar residues, *Pos* represents a positive residue, *Neg* represents a negative residue, and two types of "other" residues that do not fit in any of the previous categories are represented by *X1* and *X2*. Note that all energies in our tables have qualitative values. In d) *H* represents hydrophobic and *P* represents polar residues.

In the next step of the algorithm, the diffuser increases the probability of answer states, marked by the oracle, in the circuit (Fig. 1-A). In the final step, the results of the circuit are measured. To increase the probability of finding the answer states, the oracle and the diffuser steps should be repeated for the $\mathcal{O}(\sqrt{N/M})$ number of iterations (*R*), where *M* is the number of answer states in the circuit[22].

Three main oracle models are developed in this study, each with a distinct protein structure representation, distance dependencies, and pair-wise interaction energy tables. In the first model, which is addressed as the "SP" (simplified) model, there are no distance dependencies between the designable sites, and the 2D lattice model-like structures represent the protein (Fig. 2-a, b and c). Moreover, only integer numbers are used for energies (Fig. 3-a), and the oracle only uses the summation function. In the second model, i.e., the "MR" (more realistic) model, the pair-wise energy table (Fig. 3-b) and pre-computed reciprocals of the distances ($d^{-1}$) between the designable sites are introduced in the circuit (Fig. 2-d and e). First, the distance reciprocals are multiplied by the pair-wise energy of residues filling each designable site. Then, the values are summed to find the system's total energy. In this model, fixed-point decimal numbers are used to calculate the results (Fig. 3-b). For the SP and MR models, since we are in the noisy intermediate-scale quantum (NISQ) devices era [26], we use quantum computer simulators to study the circuits' validity and results. However, to test the practicality of our algorithms on real IBM quantum devices, a third model is developed, which is a simplified version of the SP model, addressed as "IBM-SP". This model uses the hydrophobic-polar energy table in Fig. 3-d and the protein structure in Fig. 2-a.

Note that in the MR model, the $d^{-1}$ mainly acts as weights of interactions, providing a weighted impact of each pair-wise interaction in the system. However, considering $E_{coulomb} = \frac{kq_i q_j}{d}$, where $k$ is the Coulomb constant, $q$ is the electric charges, and $i$ and $j$ are the two interacting particles, one can treat the values in Fig. 3-b as constants representing $kq_i q_j$ between residues $i$ and $j$. Thus, the MR model is directly implementing the Coulomb potential for protein design in its current form. Moreover, since we feed the pre-computed values/tables to our algorithms, the MR model can implement all potential energies for protein design in conventional packages, such as Rosetta biomolecular modelling suite[27]. For example, one can input the cosine value of distances or angles for the bonded or angular interactions.

Our results show that by using the quantum simulators, our circuits offer the expected $\mathcal{O}(\sqrt{N/M})$ queries to find the answers states, which confirm the utilization of Grover's algorithm's properties in them. However, the results of real quantum computers indicate the need for devices with much lower noise to implement our circuits.

## 2. Results

### 2.1. Number of qubits in the circuit

The number of qubits required to represent a residue in a unique binary format is given by $g = \lceil log_2(total\ number\ of\ residues) \rceil$. Thus, g=3 for the eight residues used in both SP and MR models (Fig. 3-a, b and c). Even though our energy tables with eight residues are simpler than the canonical model with 20 residues (g=5), they are still more complicated than the widely used HP table in the IBM-SP model (g=1).

Table 1 shows the number of qubits required by different circuits in our study. The total number of n qubits in the circuits is given by

$$n = g \times s \text{ (Eq. 1)},$$

where $g = \log_2(number\ of\ residues)$. For the SP and MR models, $n = 3 \times s$, while this number for the IBM-SP model is $n = 1 \times s$.

In addition, we require m number of qubits in the circuit set to represent each numeric value, e.g., the energy of each pair-wise interaction, as a part of work qubits (Fig. 1). The required number of qubits to represent a numeric value can be found as

$$m = \lceil log_2((|E_{max}| + |E_{min}|) \times i \times 2) \rceil + p \text{ (Eq. 2)},$$

where *i* is the number of interactions in the system, *p* is the number of qubits allocated to represent the values after the decimal point, and $E_{min}$ and $E_{max}$ are the minimum and maximum values in the energy table in Fig. 3. In the SP model, *p=0* and *i* is represented as input to the circuit (Fig. 2-a). However, in the MR model

$$i = i_{max} = (\frac{s \times (s-1)}{2})  \text{ (Eq. 3)}.$$

Here, since we use fixed-point decimal numbers, setting *p=5* in Eq.2 provides the precision of 0.03125 that is the default for MR model. We also study this model with more precision (MR-MP), where *p=10* and the circuit can represent values smaller than ~0.001. More detail on how to choose the minimum required number of *m* qubits for each system is provided in the Supplementary section S-I.

As shown in Table 1, the number of work qubits is $2m + 1$ and $15m + 1$ for the SP and MR models, respectively. Since we use different calculations in the oracle, the required number of work qubits differs for the two models (discussed in the Methods section).

Table 1: A brief description of each circuit in this study.

| Circuit (model) | n | N | m | Oracle's work qubits | Total qubits (q) |
|---|---|---|---|---|---|
| s=2 (SP) | 6 | 64 | 4 | 9 | 15 |
| s=3 (SP) | 9 | 512 | 5 | 11 | 20 |
| s=6,i=3 (SP) | 18 | 262,144 | 6 | 13 | 31 |
| s=6,i=4 (SP) | 18 | 262,144 | 6 | 13 | 31 |
| s=6,i=5 (SP) | 18 | 262,144 | 7 | 15 | 33 |
| s=2 (MR) | 6 | 64 | 9 | 145 | 151 |
| s=2 (MR-MP) | 6 | 64 | 14 | 225 | 231 |
| s=3 (MR) | 9 | 512 | 9 | 145 | 154 |
| s=3 (MR-MP) | 9 | 512 | 14 | 225 | 234 |
| s=2 (IBM-SP) | 2 | 4 | 2 | 5 | 7 |

From Eq.1, Eq.2, and Eq.3, we have that the maximum of total number of qubits (*q*) grows as $\sim(3 \times s + c_1 log_2(s) + c_2)$, where *c₁* and *c₂* are constant values. Trying to calculate the distance reciprocals on the same quantum circuit, as discussed by Bhaskar *et al.*[28], would add $\mathcal{O}(s^2 - c_3 \times s)$ qubits to the circuit, where *c₃* is a positive constant value.

Moreover, since the number of qubits required for the SP and MR model is large, we use the matrix product state (MPS) simulator[29], the only simulator that currently can simulate circuits with this many qubit. More detail is provided in the Methods section.

## 2.2. Finding the Answer States

Figure 4 shows the probability of finding every individual state for four different systems containing two designable sites in the SP and MR models. Here, the answer states are clearly distinguished with their higher probabilities over the other states in the system. The answer states for all systems in this study are provided in detail in Supplementary section S-II.

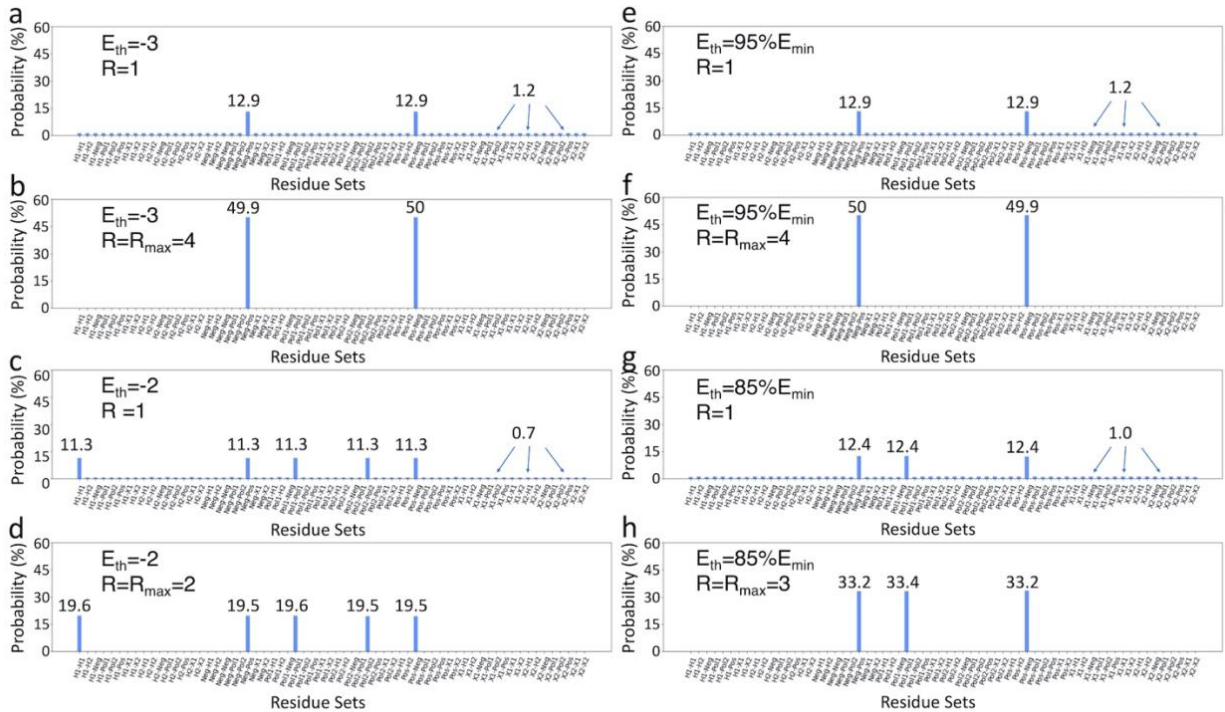

Figure 4: Histogram representations of the probability of finding each state (64 in total) in circuits with two designable sites.
Results for circuits in the SP model with: a) $E_{th}$=-3 and $R$=1; b) $E_{th}$=-3 and $R=R_{max}$=4; c) $E_{th}$=-2 and $R$=1; d) $E_{th}$=-2 and $R=R_{max}$=2. Results for circuits in the MR model with: e) $E_{th}$=95%$E_{min}$ and $R$=1; f) $E_{th}$=95%$E_{min}$ and $R=R_{max}$=4; g) $E_{th}$=85%$E_{min}$ and $R$=1; h) $E_{th}$=85%$E_{min}$ and $R=R_{max}$=3.

In the SP model with two designable sites, by setting the $E_{th}$ to −3 (using Eq. 6 in the Methods section), the algorithm finds the lowest number of answer states. Here, the algorithm finds two distinct results: *Pos-Neg* (i.e., residue 1 is *Pos* and residue 2 is *Neg*) and *Neg-Pos*, as shown in Fig. 4-a and b. These two results are expected as they are the lowest values in the pair-wise energy table with $E=-4$ in Fig. 3-a. Increasing $E_{th}$ to $-2$, the circuit finds five answer states, adding three new states to the previous two answer states from the $E_{th}=-3$ (Fig. 4-c and d).

Unlike the SP model, in the MR model, since we use decimal numbers, we can choose the $E_{th}$ value more precisely (Eq. 7 in the Methods section). To choose the answer states within the *5%* and *15%* range of the minimum energy of the system, the $E_{th}$ is set to *95%$E_{min}$* and *85%$E_{min}$*, respectively. Here, the $E_{th}=95\%E_{min}$ leads to two answer states, while choosing the $E_{th}=85\%E_{min}$ provides three answer states (Fig. 4-e, f, g, and h).

The results in Fig. 4 show that even with a single iteration ($R=1$), finding each answer state is more probable than finding each non-answer state for $s=2$ systems. However, the total probability of finding the answer states is less than the total probability of finding the non-answer states at $R=1$, which is generally correct unless for systems with the maximum number of iterations ($R_{max}$) value of 1 or 2. In other words, to have a higher probability of picking the answer states (no matter which one) among all N possible states, the total probability of finding the answer states should be higher than *50%*. This probability increases to *~100%* by increasing the number of iterations to $R=R_{max}$ (Eq. 8 in the Methods section). For example, the total probability of answer states in the *$s=2,E_{th}=-3$* system in the SP model is only *~26%* (*~12.9%* each state), while the probability of finding non-answer states is *~74%* (Fig. 4-a). Nevertheless, using $R=R_{max}=4$, the probability of finding the answer states increases to *~100%*, and the probability of all other states becomes almost zero (Fig. 4-b).

It should be noted that the results for the $s=2, E_{th}=-3$ system in the SP model (Fig. 4-a and b) are almost identical to the results of the $s=2, E_{th}=95\%E_{min}$ system in the MR model (Fig. 4-e and f). Regardless of the oracle complexity in each model, since these two systems have the same number of answer states (M=2) out of the same number of total states (N=64), the probability of finding each state is the same for both systems. Moreover, since the *M* and the *N* are the same for these systems, they have the same behaviour with changing the number of iterations, which will be discussed in more detail later in this section.

In the case of systems with $s=2$, the only recognizable difference between the SP and the MR models is the higher accuracy in choosing the $E_{th}$. Here, the role of the $d^{-1}$ in the MR model is suppressed due to the spatial symmetry in the two designable site systems. However, the $s=3$ system with different distances between each site (Fig. 2-e) breaks the symmetry and illustrates the effect of $d^{-1}$ in the MR model. To show this effect, we compare the results for the structure shown in Fig. 2-e with a similar system with complete spatial symmetry, i.e., an equilateral triangle, where the $d_{ij}=1$. The same answer states are provided for the $s=3, E_{th}=70\%E_{min}$ *system* with the symmetrical and asymmetrical configurations. The same is correct for the $s=3, E_{th}=80\%E_{min}$ *system.* Nevertheless, by choosing the $E_{th}=50\%E_{min}$, the system with the symmetrical configuration has 10 answer states while the asymmetrical system produces 19 answer states (results provided in Supplementary section S-II).

### 2.3. Role of Number of Iterations

Figure 5 shows the probability of finding the answer states as a function of the normalized number of iterations, $R/R_{max}$. Here, the probability curves represent a universal pattern for different systems in the SP and the MR models, whereby increasing the *R*, the probability of finding the answer states increases, reaching ∼*100%* at *R=R$_{max}$*. Since the computational cost of simulating larger systems is high, only the first few iterations are simulated for *s=6* systems in the SP model (Fig. 5-a), while running them for *R=R$_{max}$* would require years of simulation on CPU and terabytes of RAM (discussed in Supplementary sections S-III in detail). Moreover, the probability curves in Fig. 5 follow the $\sim sin^2(\alpha R)$, which is expected behaviour of Grover's algorithm[22,30], where $\alpha$ is a constant value (more detail in the Supplementary sections S-IV).

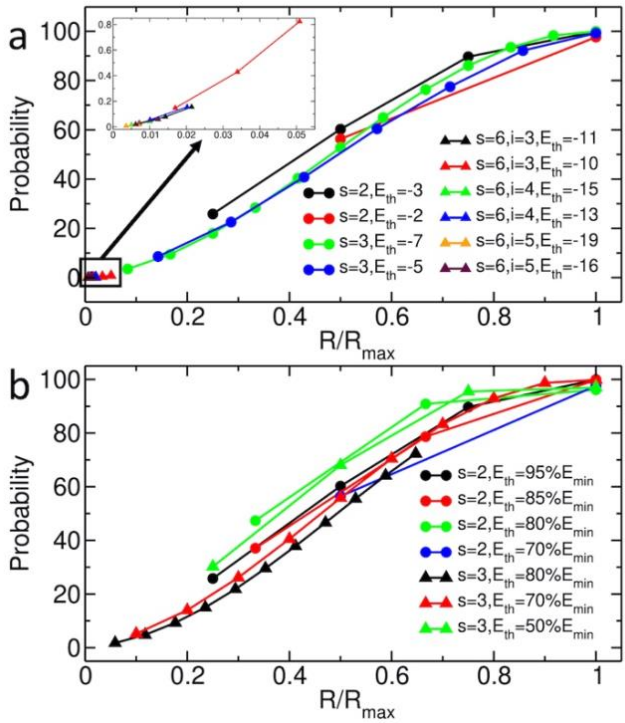

Figure 5: Probability of finding answer states in different systems as a function of *R/R$_{max}$*.
Results for a) The SP model; b) The MR model; The inset in a), represents data for the first few *R* for systems with six designable sites.

The *R$_{max}$* values obtained from simulations of systems in both SP and MR models, i.e., the ones reached to the *R/R$_{max}$=1* (Fig. 5), are plotted against *N/M* in Fig. 6. These results show that the circuits follow the $R_{max} \propto \sqrt{N/M}$ behaviour, the quantum advantage that Grover's algorithm is expected to provide.

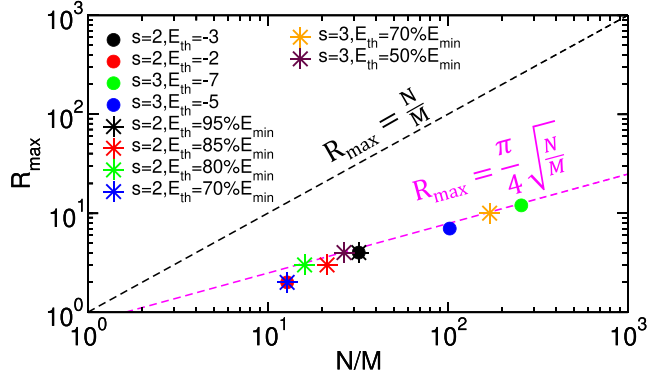

Figure 6: The $R_{max}$ values as a function of *N/M*. The circles represent the data for the SP model and the stars are the data for the MR model.
The magenta dashed line shows the $R_{max}$ threshold for Grover's algorithm, while the black dashed line is the threshold of the classic realm.

Moreover, the results for the *SP,s=2,$E_{th}$=–3* system and the *MR,s=2,$E_{th}$=95%$E_{min}$* system show that since *N/M* values are the same, the $R_{max}$ values are identical (Fig. 6). The same is correct for the *SP,s=2,$E_{th}$=–2* and the *MR,s=2,$E_{th}$=70%$E_{min}$* systems.

## 2.4. Number of gates used for classic and quantum algorithms

To compare conventional classical methods with our quantum circuits, assume we have a classical algorithm designed to search through all possible states to find the answers. Furthermore, suppose the same input data we use in quantum circuits are provided for the classic models, i.e., protein structures, interaction patterns, pre-computed pair-wise energy tables and distance reciprocals. Also, assume that the same number of bits and qubits are allocated to represent a value in both classic and quantum algorithms, i.e., *m* is the same. The latter confirms that the number of computations and the accuracy of the calculated values are similar. Despite longer bits/qubits providing higher accuracy, they require more computations (gates) units. Note that even though the total number of qubits in our quantum circuits is limited to *q* (Table 1), the total number of bits in the classical approach is not limited.

As discussed earlier in this paper, a classical search algorithm requires $\mathcal{O}(N)$ iterations to find the answer states. In contrast, our quantum circuits only require $\mathcal{O}(\sqrt{N})$ iterations. Nevertheless, this comparison does not consider the number of computations used in the classic and quantum approaches. In our quantum algorithms, the total number of computations is given by

$$\#ofQ_{tot} = \#ofQ_{init.} + \mathcal{O}(\sqrt{N}) \times (\#ofQ_{orcl.} + \#ofQ_{diff.}) \text{ (Eq. 4)},$$

where $\#ofQ_{init.}$, $\#ofQ_{orcl.}$, and $\#ofQ_{diff.}$ are the number of computations conducted in the initialization step, the oracle and the diffuser, respectively (Fig 1-a). However, in the classic realm, the total number of computations ($\#ofC_{tot}$) required to go through all possible states to find the answers is

$$\#ofC_{tot} = \mathcal{O}(N) \times (\#ofC) \text{ (Eq. 5)},$$

where $\#ofC$ represents all computations required to find the energy of a single state. For simplicity, we compare the SP model with a similarly complex classic model, referred to as "SP-classic". In the SP-classic model, similar to the SP model, the pair-wise energies of designable sites are added together for each *N* combination of amino acids (sequence) and subtracted from the $E_{th}$ to find the answer states. In the classic computation, once the addition between two numbers occurs in the SP-classic, the bits get restored and ready for the next number. However, for the SP model, due to the quantum nature of the circuit, the qubits that will be re-used should be "cleaned" by re-applying the gates (i.e., the computation is doubled compared to the SP-classic case). Moreover, the oracle in the quantum circuit should be cleaned, meaning all the gates should be re-applied (Fig. 1-b).

Details of calculating the number of computations for each step of circuits in the SP and SP-classic models are provided in Supplementary section S-V. For the SP model, the number of computations in the initialization step is $\#ofQ_{init.} \sim \mathcal{O}(\log_2(N))$ gates. The oracle cost ($\#ofQ_{orcl.}$) can be broken into the cost of introducing the energies, cost of adders, cost of subtracting the $E_{tot}$ from the $E_{th}$, and the negation cost. For introducing the energies, the total cost changes as $\sim \mathcal{O}\left[(\log_2(\log_2(N)) \times (\log_2(N))^2)\right]$. Moreover, it is shown that the number of gates required in adder functions is $\mathcal{O}(m)$[31,32], requiring $\sim \mathcal{O}[\log_2(\log_2(N)) \times ((\log_2(N))^2]$ gates to compute the $E_{tot}$. Furthermore, the cost of subtracting the $E_{tot}$ from the $E_{th}$ is $\sim \mathcal{O}[\log_2(\log_2(N))]$, and the negation cost is $\sim \mathcal{O}(1)$. Thus, $\#ofQ_{orcl.} \sim \mathcal{O}[\log_2(\log_2(N)) \times \{(\log_2(N))^2 + \log_2(N) + 1\}]$. Finally, in the diffuser, the number of computations is proportional to $\sim \mathcal{O}(\log_2(N))$. Thus, using Eq. 4 to find the total computation cost of computation, we have $\#ofQ_{tot} \sim \mathcal{O}\left[(\log_2(N)) + \sqrt{N} \times \{[\log_2(\log_2(N)) \times ((\log_2(N))^2 + \log_2(N) + 1)] + \log_2(N)\}\right]$.

For the SP-classic model, we ignore the cost of introducing the energies to the classic circuits due to the lack of information on this part. Thus, we only consider the cost of adding the energy values to find the $E_{tot}$ and subtracting it from $E_{th}$ to find the answer states. In this case, the cost of the computation for the SP-classic model from Eq. 5 is $\#ofC_{tot} \sim \mathcal{O}(N \times [\log_2(\log_2(N)) \times (\log_2(N) + 1)])$.

These estimations show that for *N>56*, the number of computations in classic circuits is larger than the quantum algorithm, despite ignoring the cost of introducing the energy values to the classic circuit. Note that for the smallest case of *s=2*, *N* is 64, indicating that the number of computations in classic form is higher than the quantum circuit for all systems. Similar is correct for the MR model, which is discussed in more detail in Supplementary section S-V.

**2.5. Real Devices and the Effect of Noise in Simulators**

The IBM-SP model circuit of our algorithms implemented on real quantum devices has four possible answer states, i.e., *HH*, *HP*, *PH* and *PP*. As expected from the energy table (Fig. 3-d), by setting the $E_{th}=0$ and using the ideal noise-free QASM[33] simulator, the circuit finds the HH state as the answer (Fig. 8-a).

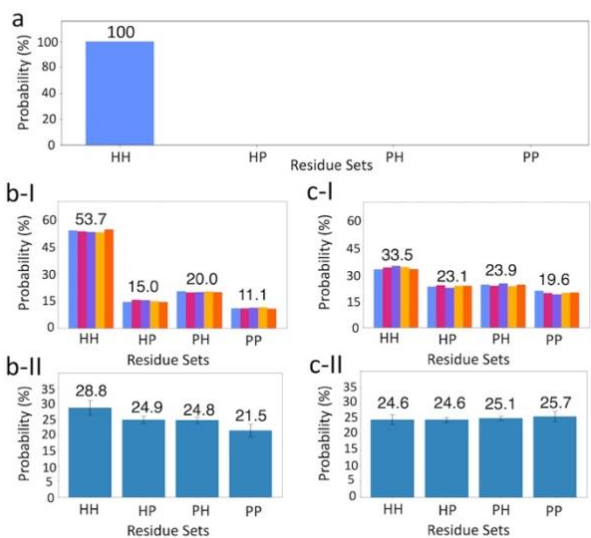

Figure 7: Histogram representations of probabilities for each state in the IBM-SP model circuit.
a) Results of the circuit, using the ideal MPS quantum computer simulator. b) Results of the circuit using: I) gate fidelity, measurement fidelity, initialization fidelity and qubit mapping of *ibmq_toronto* in the simulator; II) Real *ibm_toronto* device averaged over 20 different runs. c) Results of the circuit using: I) gate fidelity, measurement fidelity, initialization fidelity and qubit mapping of *ibmq_montreal* in the simulator; II) Real *ibm_montreal* device averaged over 30 different runs. In b) and c), the data bars with different colours show results for five separate runs. The error bars represent the standard deviation from the mean value. The numbers on each set of bars show the average probability of the state. The number of samplings for all plots is set to 8,192 shots.

In addition to the ideal simulations, we implement selective noise properties of the *ibmq_toronto* (Fig. 8-b-I) and the *ibmq_montreal* (Fig. 8-c-I) devices in the QASM simulator. Detail is provided in the Methods section. Implementing the gate, measurement and initialization fidelities, as well as the qubit connection mapping properties of the *ibmq_toronto* and the *ibmq_montreal* devices show that the simulations predict the expected answer state with the average probabilities of ~53.7% and ~33.5%, respectively. These results show that by only considering the selected noise sources, the real quantum devices are expected to distinguish the answer state from the others. Moreover, since the *ibmq_toronto* and the *ibmq_montreal* have the same qubit connection mappings, the large difference between the predicted results in Fig. 8-b-I and c-I implies lower gate, measurement and initialization fidelities for the *ibmq_toronto* compared to the *ibmq_montreal* quantum computer. These results suggest that despite having lower quantum volume (QV)[34], the *ibmq_toronto* device is more likely to distinguish the answer state clearly.

As expected from the results of the noise-included simulations, running the circuit on the *ibmq_toronto* device provides the answer state with the highest average probability of ~28.8%, while the *ibmq_montreal* device provides almost the same probabilities for all four states (Fig. 8-b-II and c-II). The probability of finding the *HH* state using the *ibmq_toronto* device is 28.77$\pm$2.39%, while for the second probable state, i.e., *HP*, this value is 24.94$\pm$1.26. These results indicate that the *HH* state has a higher probability considering the standard deviations. Note that despite having a higher probability, the probability of finding the answer states is only ~28.8%, while the probability of finding non-answer states is ~71.2%.

The difference between the results of the simulations considering the gates fidelities and the qubit mappings (Fig. 8-b-I and c-I) and the real quantum devices (Fig. 8-b-II and c-II) show the role of the performance parameters, such as coherence, crosstalk and spectator errors on the probability of the states. In the case of the *ibmq_toronto*, these performance parameters cause a significant loss in the probability of finding the answer state. These results are in agreement with the expected results of quantum devices in the NISQ era. As Chia *et al.* [35] have discussed, circuits that could run on currently available real quantum devices are mostly limited to the circuit depth of ~20. This study suggests that the depth of our circuit transpiled on real quantum devices (i.e., ~160) is much larger than the current limit of the NISQ era quantum computers. Thus, even though the noise-included simulators predict the answer states with a high probability (Fig. 8-b-I and c-I), the "large" circuit depth hinders the answer state on real quantum devices (Fig. 8-b-II and c-II).

## 3. Discussions

This work studies developing gate-based circuits to address protein design problems by implementing a pure quantum computing algorithm, i.e., Grover's algorithm. Using ideal quantum computer simulators shows that our quantum circuits can find the *M* desired answer states among *N* total states for systems with different complexities. Furthermore, the results confirm that using the maximum number of iterations ($\mathcal{O}(\sqrt{N/M})$) provides the maximum probability of finding the answer states (~100%), indicating the quadratic advantage of our approach over classical methods on conventional computers in search for answer states.

Moreover, the maximum number of computations required in our quantum circuits is smaller than the least number of computations required for classical models, confirming the quantum advantage of our circuits over conventional methods. Furthermore, our results show that the MPS simulator can correctly simulate highly entangled algorithms such as Grover's algorithm, with a large number of qubits in the circuit (up to 234 qubits).

In this work, the simplified model, i.e., the SP model, implements protein configurations similar to the 2D square lattice model, using integer numbers for calculating the energies. However, due to the limitation of the CPU times and computational resources, the largest circuits are limited to systems with six designable sites, simulating a simple hair-pin protein structure (Figure 2-c-I) or complex and compact intra-protein configurations (Fig. 2-c-II and c-III). Adding more complexity to the system in the MR model, i.e., introducing the distances reciprocals and decimal numbers, limits the number of simulatable designable sites in a system to three. However, this model enables us to use more realistic energy terms in the Hamiltonian of a protein design problem on quantum computers, i.e., mimicking the coulomb energies. Nonetheless, all different types of potentials used in conventional protein design methods could similarly be implemented in our algorithm. Moreover, our approach in implementing the distances reciprocals and decimal numbers and using multiplication functions could also be employed in the protein folding studies with quantum computers currently limited to the lattice models[11–14].

At the current NISQ stage, the number of available qubits, their connectivity and the noise associated with using each gate limit studying pure quantum algorithms on real quantum devices. Moreover, since the depth of circuits is restricted to a few tens of gates due to the noise in the real quantum devices, our results show that even running a small protein design circuit on these devices does not provide a definitive result and requires further improvements in quantum computers. As an alternative approach, it would be of interest to use popular hybrid approaches in the NISQ era, e.g., the QAOA method, to study the protein design problem and compare results with those provided by circuits using Grover's algorithm in the simulations. Nevertheless, improving the quantum devices at the current pace encourages us that in the near future, our circuits can be implemented on real quantum computers and show the advantage of quantum computers in protein design problems.

## 4. Methods

**System Setup & Simulations**

Following the Grover's algorithm in Fig. 1, after the initialization, the oracle is programmed to implement energies and conduct required calculations to find and mark the answer states. The oracle has different sub-steps associated with it. First, the values in the energy tables (Fig. 3) are introduced to the oracle, based on the protein structure and the interaction pattern (Fig. 2). Implementing the pair-wise energies in the oracle is described in the Supplementary section S-VI. Next, the energy values for every pair-wise interaction in the structure are summed to find the total energy, $E_{tot}$, for each sequence of residues (i.e., each state). This part of the oracle is the only part of the algorithm with different setups for the SP and the MR models, which is discussed in detail later in this section. Then, the $E_{tot}$ is subtracted from a threshold energy value of the circuit for each state, and the ones with the negative result are the answer. The $E_{th}$ is explicitly set for each circuit, and by changing its value, the answer states change. Details on how the $E_{tot}$ and $E_{th}$ are calculated for the SP and MR models will be discussed later in this section. Finally, the oracle negates the amplitude of the answer states. After marking the answer states (i.e., the negation), the oracle un-computes all previous steps (except for the negation step) to clear the work qubits and prevent them from affecting the final results[36].

In the SP model, the oracle calculates the total energy of the state $k$ (out of $N$ states) using:

$$E_{tot-SP}(k) = \sum_{a>b} E_{a,b}(k),$$

where $E_{a,b}(k)$ is the energy value of the interaction between designable sites *a* and *b* in the structure (Fig. 2-a–c), while specific residues in the set *k* fill these sites (Fig. 3-a). Here, since there are no distance dependencies, all pair-wise interactions contribute equally to the total energy of the system. Moreover, the lowest $E_{th}$ value in the SP model is defined as:

$$E_{th-SP}(k) = (E_{min} \times i) + 1 \text{ (Eq. 6).}$$

This $E_{th-SP}$ provides the lowest number of answer states for each circuit in the SP model (and similarly in the IBM-SP model).

In the MR model, the total energy of the state k in the oracle is calculated using:

$$E_{tot-MR}(k) = \sum_{a>b} E_{a,b}(k) \times d_{a,b}^{-1},$$

where $d_{a,b}^{-1}$ is a dimensionless matrix representing the distance reciprocal between residues *a* and *b*. Thus, nearer residues have more contributions to the $E_{tot-MR}$. In the MR model, the $E_{th}$ is defined as:

$$E_{th-MR}(k) = B \times (E_{min} \times \sum_{a>b} d_{a,b}^{-1}) \text{ (Eq.7),}$$

where *B* is a unitless constant decimal number, less than 1. For simplicity, if *B* is set to 0.95 the $E_{th-MR}$ is referred to as $E_{th-MR}=95\%E_{min}$ in this paper. Note that in this work, the $E_{th-SP}$ and $E_{th-MR}$ can be distinguished by their values, i.e., being equal to an integer number and being a percentage of the $E_{min}$, respectively. Thus, in the paper, we refer to both as $E_{th}$, removing the "SP" and "MR" subscripts.

In the oracle, a version of the quantum ripple-carry adder introduced by Cuccaro *et al.* [31] is used for adding (and subtracting) the values for both SP and MR models, which requires *2m+1* qubits to add numbers, each represented with *m* qubits. This adder is also implemented as a part of the multiplication function employed in the MR model, using *15m+1* qubits to multiply the two numbers.

The role of Grover's diffuser (Fig. 1) is to act on the *n* qubits and increase the probability of answer states over all the other states in the circuit. To accomplish this, the diffuser changes the negative amplitude of the answer states (marked in the oracle) to positive and then increases the amplitude of these flipped states[22]. Note that since the total probability of all states is one, increasing the amplitude of the answer states (and thus the probability of finding them) decreases the amplitude of the non-answer states.

The final step of the algorithm is the measurement (Fig. 1), which is done on all n qubits to find the M answer states among all N possible states. Note that the work qubits are not measured and are discarded.

In Grover's algorithm, the upper bound of the number of iterations required to get the answer states with the highest probability is:

$$R_{max} \leq \left\lceil \frac{1}{2} \frac{\pi}{2 \times arcsin(\sqrt{\frac{M}{N}})} \right\rceil + \mathcal{O}(\sqrt{\frac{M}{N}}),$$

while, in the $N \gg M$ limit, the equation changes to:

$$R_{max} \leq \left\lceil \frac{\pi}{4} \sqrt{\frac{N}{M}} \right\rceil \text{ (Eq.8).}$$

We use IBM's Qiskit package[37] to generate the circuits and simulate them in this study. Better known simulators such as the QASM require large amounts of RAM for the simulations, scaling as $16 \times 2^q$, requiring at least 128 GB of RAM for the largest SP model circuit with q=33 (Table 1). However, for circuits using more than 150 qubits in the MR model, this number reaches ~ 2×10³⁴ TB of RAM. Thus, due to the unprecedented size of our systems, we use the MPS simulators as the only possible choice to simulate all circuits. We use the computational resources provided by the Cedar cluster[38] to run our circuits on the quantum computer simulator.

**Real Quantum Devices and Noise-containing Simulators**

The quantum circuits in this work are composed of several single-qubit and multi-qubit (including two or more qubits) gates. Generally, several single-qubit and multi-qubit gates with low complexities are predefined in quantum computer simulators (details vary by simulator packages). However, in the case of more complex gates such as the *CC–NOT* (Toffoli) gate, the simulators decompose them into the simpler predefined gates.

Unlike simulators, real quantum devices only support a handful of native gates. Therefore, all gates are decomposed into the native gates when simulating our circuits on real quantum computers. However, depending on the type of the device and its manufacturer, these native gates may vary.

In addition to supporting only a few native gates, using the real quantum devices has further limitations. Since we are currently in the NISQ era, the qubits and gates contain noise while transferring data. Currently, IBM quantum devices have error rates of $\mathcal{O}(10^{-2})$ and $\mathcal{O}(10^{-4})$ for the *C-NOT* and single-qubit gates, respectively[39]. Moreover, IBM quantum devices have limited connectivity between qubits[39]. Thus, swap gates are required to perform two-qubit gates between not directly connected qubits, increasing the number of gates and the quantum computational cost of the circuit.

We use the circuit depth metric to measure and compare the complexity of circuits in this study. The circuit depth represents the number of gates in the longest path in a circuit [40], i.e., from the initialization to the measurement in our systems (Fig. 1). For the circuit of two designable sites (Fig. 1-a) at $R=1$ in the SP model, the simplest system studied on quantum computer simulators in our study, the circuit depth in the decomposed stage on an ideal device with fully connected qubits is ~17,000 (more discussion is provided in Supplementary section S-VII). In the current NISQ era, it is impossible to run a circuit with such a "large" depth on a real device due to the noise introduced to the results, owing to the gate fidelities and the system decoherence for deeper circuits[26]. Note that the depth of the circuit will increase significantly if the qubits are not fully connected on a quantum computer.

The IBM-SP model, executed on real quantum devices, requires seven qubits in total with $n=2$ (Table I). By setting the $E_{th}$ to 0, the circuit will have one answer state (Fig.3 d) and the $R_{max}=1$. We employ two IBM quantum computers, the *ibmq_toronto* (v1.6.1 and v1.6.2) with the IBM Quantum Falcon r4 processors and the *ibmq_montreal* (v1.10.11) with the IBM Quantum Falcon r4 processors, both having 27 qubits and the same pattern of connectivity. To run the IBM-SP model on these devices, we use optimization level 2 and a selective pattern of qubits to transpile the ideal circuit on them. Moreover, we apply the transpilation 500 times for each run and select the circuit with the lowest depth, ranging between 158 and 167, as an input for the real quantum devices. To compare these quantum computers, we use the quantum volume, a unitless number, quantifying the largest random circuit that a quantum computer can implement successfully[34]; thus, the more, the better.

**Acknowledgments**

The authors would like to thank WestGrid (www.westgrid.ca) and Compute Canada (www.computecanada.ca) for providing computational resources for this project.

We acknowledge the use of IBM Quantum services for this work. The views expressed are those of the authors, and do not reflect the official policy or position of IBM or the IBM Quantum team.

We also would like to acknowledge CMC Microsystems for facilitating this research, specifically through their member access to the IBM Quantum Hub at Institut quantique.

# Supplementary Materials for

## Gate-based Quantum Computing for Protein Design

Mohammad Hassan Khatami, Udson C. Mendes, Nathan Wiebe, Philip Kim[*]

*Corresponding author. Email: pi@kimlab.org

### I. Number of required qubits in the circuit and the general algorithm

In this section, we describe how to estimate the m value, as well as the total number of qubits for different circuits. To do so, in each circuit, we first need to estimate the largest possible positive value and smallest possible negative value. This estimation is based on the Hamiltonian and the energy table associated with the system of study.

Each value in the circuit is represented by a set of qubits shown in Fig. S1-A. One can simply check that using n qubits, we can show $2^n$ numbers. For example, by using 6 qubits, we can show 64 numbers, i.e., 0 to 63. However, if we need to show both negative and positive numbers, the most left qubit is devoted to the sign, as shown in Fig. S1-A. Thus, we can show numbers from −32 to +31 using six qubits.

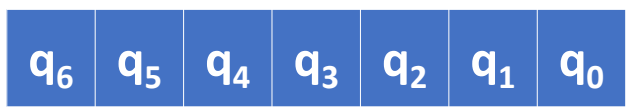

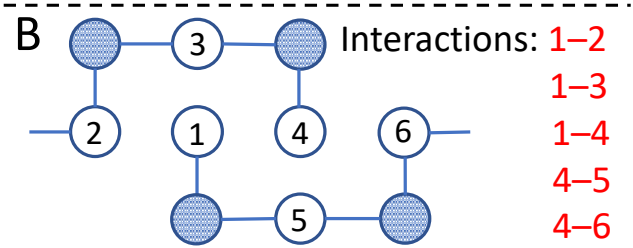

**FIG. S1**. A) Schematic representation of 7 qubits. B) Schematic representation of a system with 6 interacting residues and 5 interactions.

Here, we focus on a system in the SP model with six designable sites and five interactions, as shown in Fig. S1-B. Using the energy table in the Fig. 2-A of the main text (and Fig. S5- B), the largest energy value is +4, and the smallest is −4. Since we have five interactions, the extremes of the total energy of a set can be −20 and +20. Now, if we subtract these numbers from $E_{th}$ = *−19*, the range of values in the circuit will be from +39 to −1. Based on the calculations described earlier, using six qubits to represent numbers is insufficient for +39 to −1 range, as it can show −32 to +31 in a standard implementation.

However, using seven qubits can represent numbers in the −64 to +63 range that covers our +39 to −1 values.

In our circuits, $q = n + \#\_work\_qubits$ (Fig. S2-A). To calculating the total number of qubits required for the circuit in Fig. S1-B, we first calculate the number of *n* qubits. Since, *n = 3 × s*, and *s=6* for 6 designable sites, thus *n = 18* (Fig. S2-B).

The $\#\_work\_qubits$, which is $2m + 1$, is describe as following:
1. We allocate four qubits sets, one with 18 (generally $3 \times s$) qubits labelled as "var_qubits", one with a single qubit (required by the adder) labelled as "o_qubit" and two sets, each with 7 (generally *m*) qubits, labelled as "a_qubits" and "b_qubits".
2. The energy of the interaction 1 – 2 is put in the a_qubits.
3. The energy of the interaction 1 – 3 is put in the b_qubits.
4. The value in the a_qubits is added to b_qubits and put into a_qubits (using quantum ripple adder).
5. The energy of the interaction 1 – 3 is cleaned from the b_qubits (to inter the next value).
6. The energy of the interaction 1 – 4 is put in the b_qubits.
7. The value in the a_qubits is added to b_qubits and put into a_qubits (now we have 1-2, 1-3 and 1-4 adde together).
8. The energy of the interaction 1 – 4 is cleaned from the b_qubits.
9. The energy of the interaction 4 – 5 is put in the b_qubits.
10. The value in the a_qubits is added to b_qubits and put into a_qubits.
11. The energy of the interaction 4 – 5 is cleaned from the b_qubits.
12. The energy of the interaction 4 – 6 is put in the b_qubits.
13. The value in the a_qubits is added to b qubits and put into a_qubits.

Now, we have added all energies in the configuration. We only need 18 qubits for n and 7 + 7 qubits for the a_qubits + b_qubits and 1 qubit for the o_qubit, in total 33 qubits.

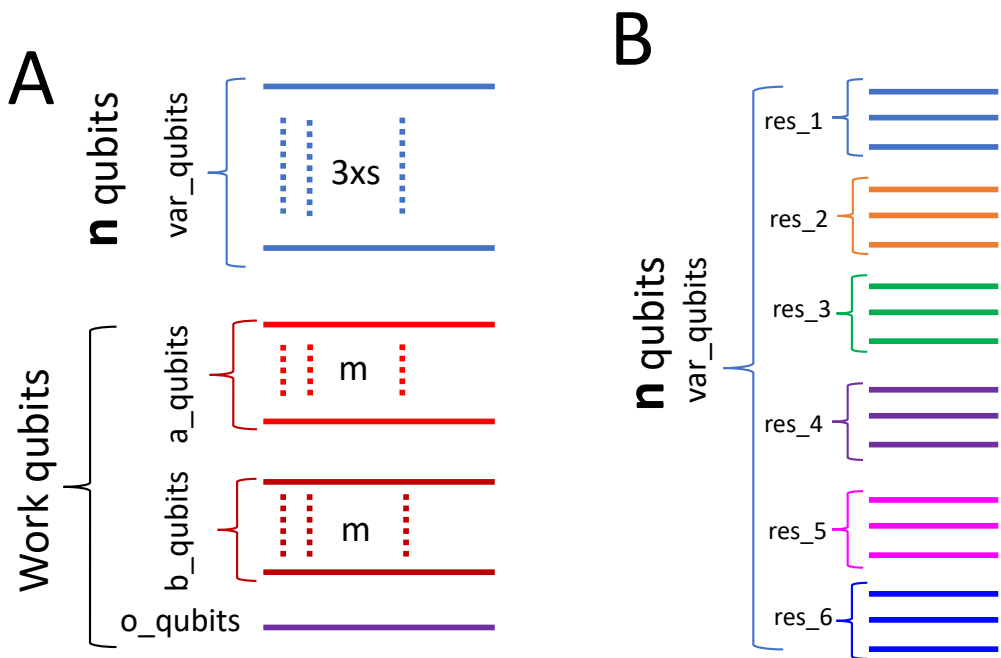

**FIG. S2**. General representation of qubits in our algorithms. A) qubits representation in the whole circuit. B) $n$ qubits for 6 designable site/residues system.

## II. Explicit results of each system

Answer states of all systems studied in the main paper are provided in Tables S1, S2 and S3.

**TABLE S1.** Answers for all systems studied in the SP model.

| s=2 | | s=3 | | s=6,i=3 | |
|---|---|---|---|---|---|
| $E_{th}$=-3 | $E_{th}$=-2 | $E_{th}$=-7 | $E_{th}$=-5 | $E_{th}$=-11 | $E_{th}$=-10 |
| Seq(2 – 1) | | Seq(3 – 2 – 1) | | Seq(6 – 5 – 4 – 3 – 2 – 1) | |
| Pos-Neg<br>Neg-Pos | H1-H1<br>Pol1-Pol1<br>Pol2-Pol2<br>Pos-Neg<br>Neg-Pos | Pos-Pos-Neg<br>Neg-Neg-Pos | H1-H1-H1<br>Pos-Pos-Neg<br>Neg-Neg-Pos<br>Pol1-Pol1-Pol1<br>Pol2-Pol2-Pol2 | Pos-Neg-Pos-Neg-Pos-Neg<br>Pos-Neg-Pos-Neg-Neg-Pos<br>Pos-Neg-Neg-Pos-Pos-Neg<br>Pos-Neg-Neg-Pos-Neg-Pos<br>Neg-Pos-Pos-Neg-Pos-Neg<br>Neg-Pos-Pos-Neg-Neg-Pos<br>Neg-Pos-Neg-Pos-Pos-Neg<br>Neg-Pos-Neg-Pos-Neg-Pos | Pol1-Pol1-Pos-Neg-Pos-Neg<br>Pol1-Pol1-Pos-Neg-Neg-Pos<br>Pol1-Pol1-Neg-Pos-Pos-Neg<br>Pol1-Pol1-Neg-Pos-Neg-Pos<br>Pol2-Pol2-Pos-Neg-Pos-Neg<br>Pol2-Pol2-Pos-Neg-Neg-Pos<br>Pol2-Pol2-Neg-Pos-Pos-Neg<br>Pol2-Pol2-Neg-Pos-Neg-Pos<br>Pos-Neg-H1-H1-Pos-Neg<br>Pos-Neg-H1-H1-Neg-Pos<br>Pos-Neg-Pol1-Pol1-Pos-Neg<br>Pos-Neg-Pol1-Pol1-Neg-Pos<br>Pos-Neg-Pol2-Pol2-Pos-Neg<br>Pos-Neg-Pol2-Pol2-Neg-Pos<br>Pos-Neg-Pos-Neg-H1-H1<br>Pos-Neg-Pos-Neg-Pol1-Pol1<br>Pos-Neg-Pos-Neg-Pol2-Pol2<br>Pos-Neg-Pos-Neg-Pos-Neg<br>Pos-Neg-Pos-Neg-Neg-Pos<br>Pos-Neg-Neg-Pos-H1-H1<br>Pos-Neg-Neg-Pos-Pol1-Pol1<br>Pos-Neg-Neg-Pos-Pol2-Pol2 | Pos-Neg-Neg-Pos-Pos-Neg<br>Pos-Neg-Neg-Pos-Neg-Pos<br>Neg-Pos-H1-H1-Pos-Neg<br>Neg-Pos-H1-H1-Neg-Pos<br>Neg-Pos-Pol1-Pol1-Pos-Neg<br>Neg-Pos-Pol1-Pol1-Neg-Pos<br>Neg-Pos-Pol2-Pol2-Pos-Neg<br>Neg-Pos-Pol2-Pol2-Neg-Pos<br>Neg-Pos-Pos-Neg-H1-H1<br>Neg-Pos-Pos-Neg-Pol1-Pol1<br>Neg-Pos-Pos-Neg-Pol2-Pol2<br>Neg-Pos-Pos-Neg-Pos-Neg<br>Neg-Pos-Pos-Neg-Neg-Pos<br>Neg-Pos-Neg-Pos-H1-H1<br>Neg-Pos-Neg-Pos-Pol1-Pol1<br>Neg-Pos-Neg-Pos-Pol2-Pol2<br>Neg-Pos-Pos-Neg-Pos-Neg<br>Neg-Pos-Pos-Neg-Neg-Pos<br>H1-H1-Pos-Neg-Pos-Neg<br>H1-H1-Pos-Neg-Neg-Pos<br>H1-H1-Neg-Pos-Pos-Neg<br>H1-H1-Neg-Pos-Neg-Pos |

| s=6,i=4 | | | s=6,i=5 | |
|---|---|---|---|---|
| $E_{th}$=-15 | $E_{th}$=-13 | | $E_{th}$=-19 | $E_{th}$=-16 |
| Seq(6 – 5 – 4 – 3 – 2 – 1) | | | Seq(6 – 5 – 4 – 3 – 2 – 1) | |
| Pos-Pos-Neg-Pos-Pos-Neg<br>Pos-Pos-Neg-Neg-Neg-Pos<br>Neg-Neg-Pos-Pos-Pos-Neg<br>Neg-Neg-Pos-Neg-Neg-Pos | H1-H1-H1-Pos-Pos-Neg<br>Pol1-Pol1-Pol1-Pos-Pos-Neg<br>Pol1-Pol1-Pol1-Neg-Neg-Pos<br>H1-H1-H1-Neg-Neg-Pos<br>Pol2-Pol2-Pol2-Pos-Pos-Neg<br>Pol2-Pol2-Pol2-Neg-Neg-Pos<br>Pos-Pos-Neg-H1-H1-H1<br>Pos-Pos-Neg-Pol1-Pol1-Pol1 | Pos-Pos-Neg-Pol2-Pol2-Pol2<br>Pos-Pos-Neg-Pos-Pos-Neg<br>Pos-Pos-Neg-Neg-Neg-Pos<br>Neg-Neg-Pos-H1-H1-H1<br>Neg-Neg-Pos-Pol1-Pol1-Pol1<br>Neg-Neg-Pos-Pol2-Pol2-Pol2<br>Neg-Neg-Pos-Pos-Pos-Neg<br>Neg-Neg-Pos-Neg-Neg-Pos | Pos-Pos-Neg-Neg-Neg-Pos<br>Neg-Neg-Pos-Pos-Pos-Neg | Pol1-Neg-Pos-Pos-Pos-Neg<br>Pos-Pos-Neg-Pol1-Neg-Pos<br>Pos-Pos-Neg-Neg-Pol1-Pos<br>Pos-Pos-Neg-Neg-Neg-Pos<br>Neg-Pol1-Pos-Pos-Pos-Neg<br>Neg-Neg-Pos-Pos-Pos-Neg |

**TABLE S2**. Answers for all systems studied in the default MR model.

| | s=2 | | | | s=3 | | | s=3 (symmetric) |
|---|---|---|---|---|---|---|---|---|
| $E_{th}$=95%$E_{max}$ | $E_{th}$=85%$E_{max}$ | $E_{th}$=80%$E_{max}$ | $E_{th}$=70%$E_{max}$ | $E_{th}$=80%$E_{max}$ | $E_{th}$=70%$E_{max}$ | | $E_{th}$=50%$E_{max}$ | $E_{th}$=50%$E_{max}$ |
| Seq(2 – 1) | | | | Seq(3 – 2 – 1) | | | | |
| Pos-Neg Neg-Pos | Pol1-Pol1 Pos-Neg Neg-Pos | Pol1-Pol1 Pol2-Pol2 Pos-Neg Neg-Pos | H1-H1 Pol1-Pol1 Pol2-Pol2 Pos-Neg Neg-Pos | Pol1-Pol1-Pol1 | H1-H1-H1 Pol1-Pol1-Pol1 Pol2-Pol2-Pol2 | H1-H1-H1 H1-H1-H2 Pos-Neg-Pol1 Pos-Neg-Pos Pos-Neg-Neg Pos-Neg-X2 Neg-Pos-Pol1 Neg-Pos-Neg Neg-Pos-X2 Pol1-Pol1-Pol1 Pol1-Pol1-Pol2 Pol1-Pol1-Pos Pol1-Pol1-Neg Pol1-Pol2-Pol1 Pol1-Pol2-Pol2 Pol2-Pol1-Pol1 Pol2-Pol1-Pol2 Pol2-Pol2-Pol1 Pol2-Pol2-Pol2 | H1-H1-H1 H2-H2-H2 Pol1-Pol1-Pol1 Pol1-Pol1-Pol2 Pol1-Pol2-Pol1 Pol1-Pol2-Pol2 Pol2-Pol1-Pol1 Pol2-Pol1-Pol2 Pol2-Pol2-Pol1 Pol2-Pol2-Pol2 |

**TABLE S3**. Answers for all systems studied in the MR-MP model.

| | s=2 (MP) | | | | s=3 (MP) | |
|---|---|---|---|---|---|---|
| $E_{th}$=95%$E_{max}$ | $E_{th}$=85%$E_{max}$ | $E_{th}$=80%$E_{max}$ | $E_{th}$=70%$E_{max}$ | $E_{th}$=80%$E_{max}$ | | $E_{th}$=70%$E_{max}$ |
| Seq(2 – 1) | | | | Seq(3 – 2 – 1) | | |
| Pos-Neg Neg-Pos | Pol1-Pol1 Pos-Neg Neg-Pos | Pol1-Pol1 Pol2-Pol2 Pos-Neg Neg-Pos | H1-H1 Pol1-Pol1 Pol2-Pol2 Pos-Neg Neg-Pos | Pol1-Pol1-Pol1 Pol2-Pol2-Pol2 | | H1-H1-H1 Pol1-Pol1-Pol1 Pol2-Pol2-Pol2 |

### III. Computational costs of the simulations on conventional computers

Since we use quantum computer simulators to study our circuits, we are restricted by the limitations of the conventional computers, i.e., the amount of RAM required to simulate a system and how long the simulation will take in real-time (CPU time) to run. Fig. S3 shows the changes in the amount of RAM and CPU time required to simulate different systems as a function of $R/R_{max}$. For all data sets, increasing the number of iterations increases both RAM amount and CPU time.

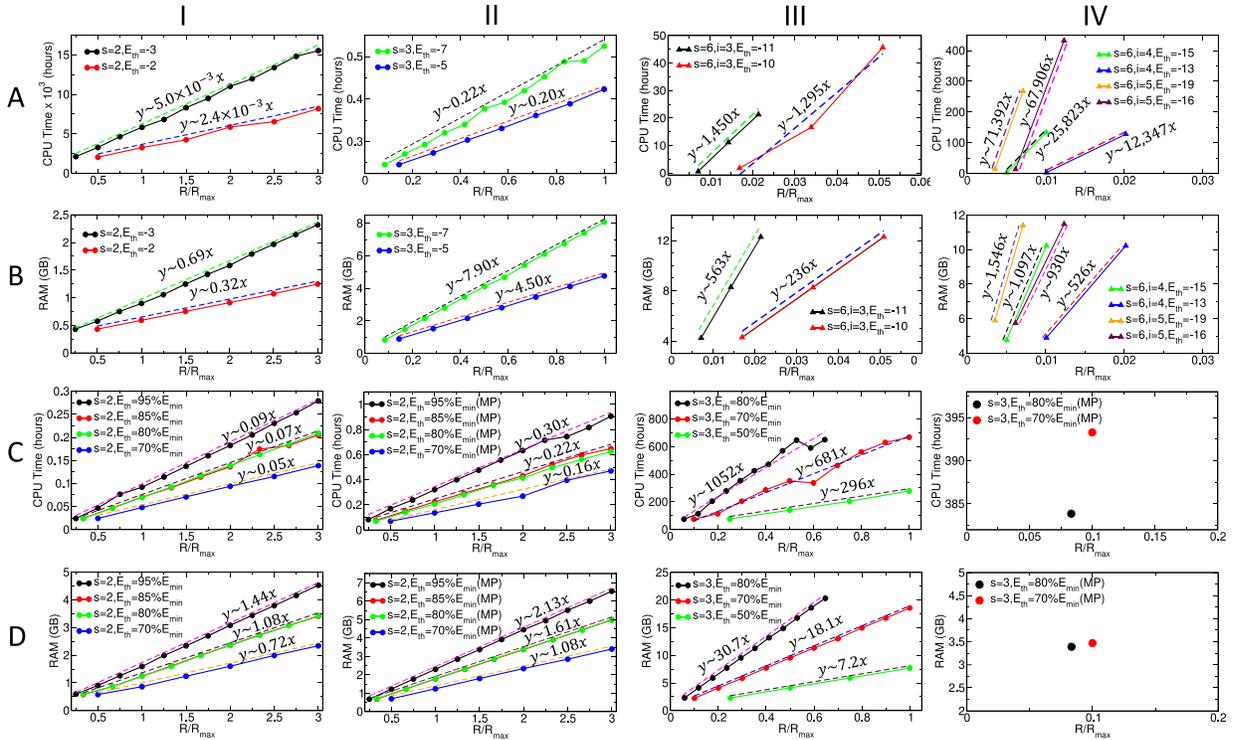

**FIG. S3**. Change in the cost of simulation as a function of normalized number of iterations for systems in the SP and MR models. Data in rows A) and B) shows the CPU time and the RAM usage for systems in the SP model, respectively. Data in rows C) and D) shows the CPU time and the RAM usage for systems in the MR model, respectively. In rows C) and D) data for the MR-MP are presented in columns II and IV. In all panels, the dashed lines represent the fit to the curves.

Based on the data from circuits with two and three designable sites in the SP model and circuits with two designable sites in the MR model (both default MR and MR-MP) represented in columns I and II in Fig. S3, we can see that the RAM amount and the CPU time are linear functions of $R/R_{max}$. This linear behaviour is expected as increasing the number of iterations increases the number of times the oracle and the diffuser are called in the algorithm. Thus, since the simulations run on conventional computational resources, increasing the amount of computation linearly with $R$ increases the cost of computation linearly. Expecting a similar linear behaviour, we predict the resources required to simulate the larger circuits that have few actual data points in columns III and IV in Fig. S3. Note that, for the systems with two designable sites in both SP and MR models, the x-axis is extended to $R/R_{max}=3$ to have more data points on the curves to fits.

These results show that circuits in the SP model with two and three designable sites, as well as the ones in the MR model with two designable sites, shown in columns I and II of Fig. S3, could run with $R = R_{max}$ on resources available on a regular laptop device as the required RAM amount is ~8 GB and the maximum CPU time is within an hour time range. However, for the rest of the systems, shown in columns III and IV of Fig. S3, even running the first few iterations takes days of simulation and more than 10 GB of RAM in some cases. In these circuits, running the simulation with $R/R_{max} = 1$ requires significant computational resources and simulation times. As an example, the $s = 6, i = 5, E_{th} = -19$ system would need ~1.5 TB of RAM and ~71,392 hours (~8.1 years) of CPU time to simulate with $R = R_{max}$ iterations (Fig. S3-A-IV and B-IV).

In the MR model (both the default and MR-MP), the computational resources required to simulate the $s = 2$, $E_{th} = 85\%E_{min}$ and the $s = 2$, $E_{th} = 80\%E_{min}$ systems are almost identical (Fig. S3-C-I and D-I, and C-II and D-II). These results motivated us to study the changes in the CPU time, and the RAM usage as a function of $R$ (instead of $R/R_{max}$) for all systems in Fig. S3, which is provided in Fig. S4 and will be discussed in more details later in this section. Based on these results, for systems with the same number of designable sites and interactions, the computational resources required to simulate the circuits are almost identical for all $E_{th}$ values, which indicates that the usage of computational resources is independent of the $E_{th}$.

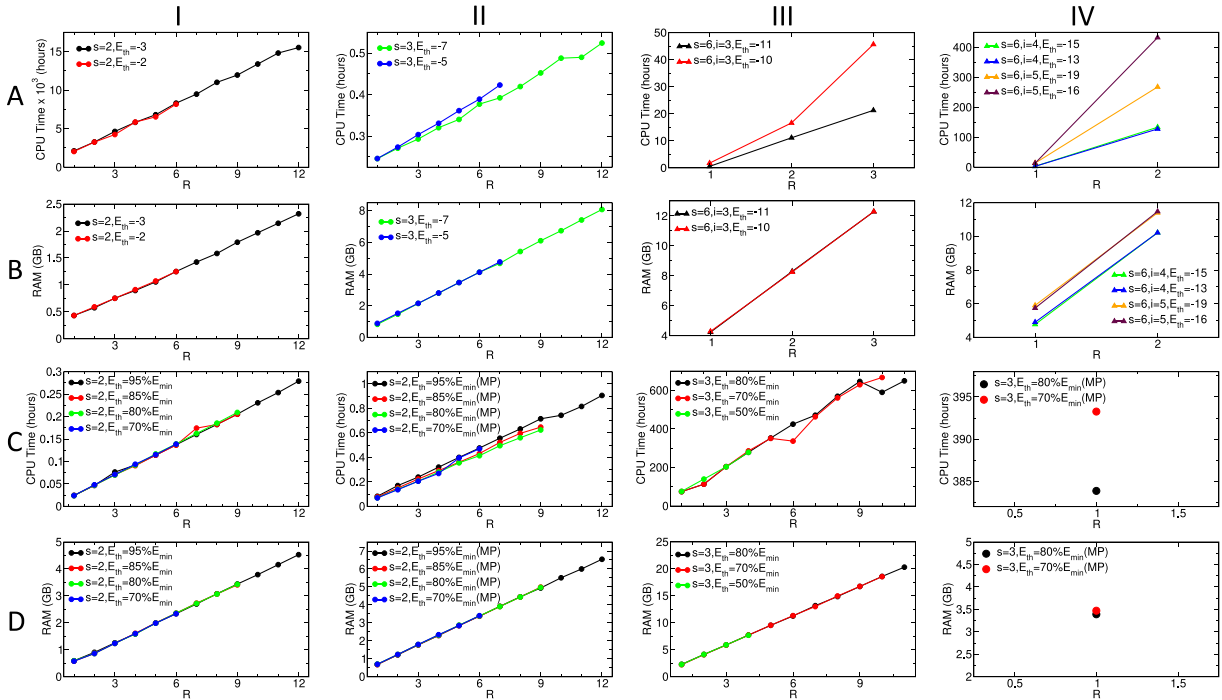

**FIG. S4**. Change in the cost of simulation as a function of R for systems in the SP and MR models. Data in rows A) and B) shows the CPU time and RAM usage for systems in the SP model, respectively. Data in rows C) and D) shows the CPU time and RAM usage for systems in the MR model, respectively. In rows C) and D) data for default MR are presented in columns I and III, and for the MR-MP in columns II and IV. Only one data point for systems in C-IV and D-IV are available in our simulations. In all panels, the dashed lines represent the fit to the curves.

Moreover, for similar systems in the SP and the MR models, i.e., the $s = 2$, $E_{th} = -3$ and $s = 2, E_{th} = 95\%E_{min}$ circuits, as well as, the $s = 2, E_{th} = -2$ and the $s = 2, E_{th} = 70\%E_{min}$ circuits, we can see that the CPU time is more than 18 times longer for the MR model systems (Figure S3-A-I and C-I). Also, the RAM usage in the MR model systems is approximately twice the ones in the SP model (Figure S3-B-I and D-I). These differences are expected as the circuits in the MR model are more complex than those in the SP model.

Similarly, comparing the results between the equivalent systems in the two designable site circuits in the MR model, i.e., the default MR and the MR-MP, show that the system with more precision requires ~3.3 times longer CPU times and ~1.5 times more RAM (Figure S3-C-I, C-II, D-I, and D-II). Thus, increasing the precision in a simulation increases the computational cost for similar systems in the MR and the MR-MP models. This is expected since to increase the precision, more qubits are needed in the circuit (Table I in the main text).

Figure S4 reports the CPU time and RAM usage of the system studied as a function of $R$. In most panels, the curves are overlapping, showing similar resource usage as a function of iteration. However, in some cases, some data points have different values, which do not fit the general trends of curves in similar data sets. This only occurs in the CPU time plots and it is due to different properties of CPUs in the Cedar cluster, which we run our jobs on it. Nevertheless, these differences in the CPU times do not affect the generality of our conclusions in the main text. In case of the $s=6, i=5, E_{th}=-19$ in the SP model, adjusting the current CPU time with the one in a faster CPU device could lead to ~36,500 hours (instead of 71,392 hours), which is around 4.5 years and still unachievable.

## IV. Role of number of iterations

Figure S5 represents the change in the probability of finding answer states as a function of the number of iterations for selected systems in the SP and MR models. We can see that the probability curves follow the $\sim sin^2((aR + b)/c)$ patterns, where a, b and c are constant values. This pattern is an expected behaviour of Grover's algorithm and is more evident and smoother for the system with three designable sites in the SP model (Fig. S5-B). By increasing the number of iterations, the probability of finding answer states increases. However, after reaching the $R_{max}$ (Eq. 8 in the main paper), which is the first peak value in all curves in Fig. S5, the probability of finding answer states decreases.

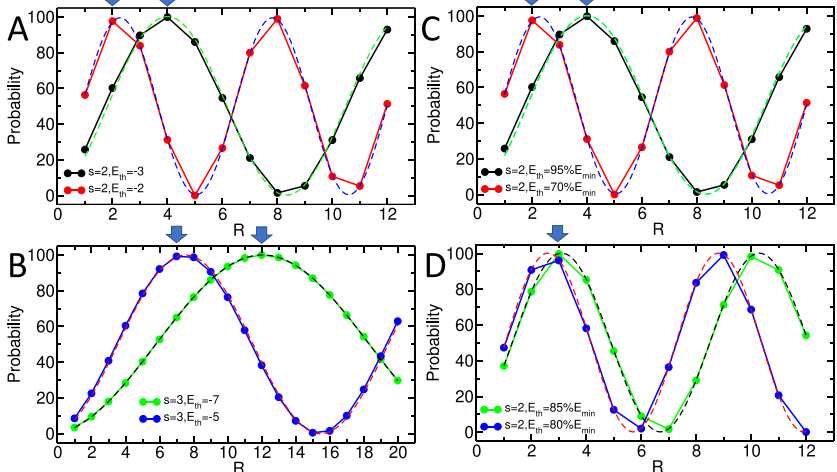

**FIG. S5**. Probability of finding answer states as a function of the number of iterations R in different systems. The SP model with: A) Two designable sites; B) Three designable sites. The MR model with two designable sites with: C) Eth = 95%Emin and Eth = 70%Emin; D) Eth = 85%Emin and Eth = 80%Emin. The dashed lines show the sin2((aR + b)/c) curves fitted to each data set. The first local maximum on each curve is pointed by the blue arrow.

Moreover, the curves for the systems with two designable sites in the SP model with $E_{th}$ = −3 and $E_{th}$ = −2 (Fig. S5-A) and the ones in the MR model with $E_{th} = 95\%E_{min}$ and $E_{th} = 70\%E_{min}$ (Fig. S5-C) are almost identical. As discussed earlier, this behaviour is expected for systems with the same total number of states $N$ and the same number of answer states $M$.

Nevertheless, two systems can have the same $R_{max}$ value, but different patterns for the probability of finding the answer states as a function of $R$. For example, for two systems in the MR model, i.e., $s = 2$, $E_{th} = 85\%E_{min}$ and $s = 2$, $E_{th} = 80\%E_{min}$ systems, with three and four answer states respectively, the $R_{max}$ value is 3 (Fig. S5-D). However, despite having the same $R_{max}$, the curves for the two systems differ significantly, where even their second peaks occur in different $R$ values. This result shows that despite having the same $N$ states and the same $R_{max}$ value, if two systems have different $N/M$, their probability curves will have different patterns.

## V. Computational costs in the circuits

### V.I: quantum circuit

To find the number of gates for each step of Grover's algorithm implemented in our circuits, deep knowledge of each step is required. To this end, we investigate the SP model circuits.

The number of computations in the initialization step ($\#of Q_{init.}$) is equivalent of number of $H$-gates used, which is $n$ (Fig. S7-C). Thus, $\#of Q_{init.} \sim \mathcal{O}(n) = \mathcal{O}(\log_2(N))$.

Next, we need to calculate the cost of computation for the oracle, i.e., $\#of Q_{orcl.}$. Here, the computation cost consists of introducing energies to the work qubit, adding them up and negating the answer states, and finally cleaning up the gates (Fig. 1-b in the main text).

To introduce energies, we use the *6-control-1-NOT* gates represented in Fig. S7-C. The maximum required *6-control-1-NOT* gates will be used to represent value of −1 in the circuit, which requires applying 7 (as *m=7*) of the *6-control-1-NOT* gates. For the other values in the pair-wise energy table, lesser number of the gates are required. However, to be on the safe side, we will assume all values in the energy table are −1. For each value, we require some *X* gates as shown in Fig. S7-C. Similarly, we consider the maximum number of *X* gates that is required, which is 12 (as $2 \times 2 \times [log_2(total\ number\ of\ residues)]$, which is a constant for our circuits) gates for the *H1-H1* pair. Thus, introduce energy values each time we require
$$8 \times 8 \times [m \times (6CX) + 12 \times X]\ \text{(Eq. S1)}$$
gates, where *6CX* represents the *6-control-1-NOT* gates. From Eq. 1, Eq. 2 and Eq. 3 in the main text, we have that $m \sim \mathcal{O}(\log_2(\log_2 N))$, which simplifies the Eq. S1 to $\sim \mathcal{O}(\log_2(\log_2 N)) \times (6CX)$. It is not possible for current quantum computers to implement the *6-control-1-NOT* gates as a single gate, which is a task for future quantum computers with fully connected qubit mappings. From the Nielsen-Chuang book on "Quantum computation and quantum information", we have that an n-*contol-1-X* gate can be decomposed into *2(n–1) CCX* gates, where each *CCX* can be decomposed into 15 single and CX gates. Thus, Eq. S1 with decomposition of *6CX* simplifies to $\sim \mathcal{O}(\log_2(\log_2 N)) \times \mathcal{O}(\log_2 N) \approx \mathcal{O}(\log_2(\log_2(N)) \times \log_2(N))$. Thus, the number of computations required for introducing the energies to the circuit is $\sim\mathcal{O}(\log_2(\log_2(N)) \times \log_2(N))$. Note that for a fully connected mapping, this number will be $\sim\mathcal{O}(\log_2(\log_2(N)))$ instead of $\sim\mathcal{O}(\log_2(\log_2(N)) \times \log_2(N))$. For the *SP,s=2* system with any $E_{th}$ value, the introducing the energy is done once (description is provided in the algorithm part of section I of the supplementary data). However, for each bond *i* in the structure, we need to introduce the energies for them. However, for *s>2* (i.e., *i>1*) we need to clean qubits from values of the previous bond (by re-applying the gates used to introduce them in the first place) and add the new bond's energy values (algorithm part of section I of the supplementary data). This requires applying the introducing the energy step $2i - 1$ times, where $i \sim \mathcal{O}(\log_2(N))$. Thus, the total cost of the introducing energies at worst case scenario will be $\sim\mathcal{O}(\log_2(\log_2(N)) \times \log_2(N)) \times \mathcal{O}(\log_2(N)) \approx \mathcal{O}[\log_2(\log_2(N)) \times (\log_2(N))^2]$.

Then, there is the cost of adders, which is discussed briefly in the main text and changes as $\sim\mathcal{O}(m) \approx \mathcal{O}(\log_2((\log_2(N))))$. Thus, to add the energies together, each time an $\sim\mathcal{O}(\log_2((\log_2(N))))$ computation is implemented in the circuit. Since energies of each interaction *i* is added to find the $E_{tot}$, the computation cost is $\sim\mathcal{O}(\log_2(\log_2(N)) \times i \approx \mathcal{O}(\log_2((\log_2(N))) \times \mathcal{O}(\log_2(N)) \approx \mathcal{O}(\log_2((\log_2(N)) \times \log_2(N))$.

After summing the energies between all designable sites, circuits subtract the $E_{tot}$ from the $E_{th}$, which adds another $\sim\mathcal{O}(\log_2(\log_2(N)))$ computation to the system (same as adder function).

For the negation, a single *CZ* gate is used for negating the answer states. Finally, the work qubits are cleaned by applying the inverse of the functions initially implemented, meaning that all computations applied before the negating process will duplicated.

Thus, the total number of computation used in the oracle is $2 \times [cost\_energy\_introduction + cost\_adders\_E_{tot} + cost\_subtract\_E_{th} + cost\_CZ\_gate]$.
Using the values provided in previous discussions in this section, the cost of the oracle will be
$\sim\mathcal{O}\left[(\log_2(\log_2(N)) \times (\log_2(N))^2) + (\log_2((\log_2(N)) \times \log_2(N)) + \log_2(\log_2(N))\right]$,
or
$$\sim\mathcal{O}[\log_2(\log_2(N)) \times ((\log_2(N))^2 + \log_2(N) + 1)]$$

Finally, we should consider the cost of computation in the diffuser step ($\#ofQ_{diff.}$). From the Nielsen-Chuang book on "Quantum computation and quantum information", we can see that the diffuser is composed of *2(n+1) H*-gates, *2n X*-gates and one *(n-1)-control-1-NOT* gate. The *H*-gates and *X*-gates each require $\sim\mathcal{O}(\log_2(N))$ computations. However, based on the previous discussions in this section, for the *(n-1)-control-1-NOT* gate we require $2((n-1)-1)CCX$ gates, which leads to $\sim\mathcal{O}(\log_2(N))$ gates. Thus, in general $\#ofQ_{diff.} \sim \mathcal{O}(\log_2(N))$.

The total cost of a quantum algorithm will be calculated as, $\#ofQ_{tot} = \#ofQ_{init.} + \sqrt{N} \times (\#ofQ_{orcl.} + \#ofQ_{diff.})$. Using the discussions in this section, the $\#ofQ_{tot}$ will require
$\sim \mathcal{O}\left[(\log_2(N)) + \sqrt{N} \times \{[\log_2(\log_2(N)) \times ((\log_2(N))^2 + \log_2(N) + 1)] + \log_2(N)\}\right]$ gates. (Fig. S6).

### V.II: classic circuit

Since we do not have any information on implementing the energies in the classical circuits, we ignore its cost in our calculations. Thus, only the adder circuit cost remains. In a classic algorithm, the circuits require adders to add the energy values provided in the energy tables for each interaction. Using a similar-to-quantum adder requires $\sim \mathcal{O}(m) \approx \mathcal{O}(\log_2(\log_2(N)))$ computations. Applying running it for each interaction *i*, requires $\sim \mathcal{O}(\log_2(N))$ computation. Thus, the total cost of an adder circuit, and in general the cost of computation for classic circuit ($\#ofC$), is $\sim \mathcal{O}(\log_2(\log_2(N)) \times \log_2(N))$, similar to quantum circuits. In addition, there is a cost of subtracting the $E_{tot}$ from the $E_{th}$, which adds another $\sim \mathcal{O}(\log_2(\log_2(N)))$.

The total number of computations for classic algorithm is $\#ofC_{tot} = \mathcal{O}(N) \times (\#ofC)$, which is simplified as $\sim \mathcal{O}(N \times [(\log_2(\log_2(N)) \times \log_2(N)) + \log_2(\log_2(N))])$, which could be simplified as $\sim \mathcal{O}(N \times [\log_2(\log_2(N)) \times (\log_2(N) + 1)])$ (Fig. S6)

### V.III: comparing quantum and classic

Based on the data provided in Fig. S6, we can see that from the small N values (i.e., N>56), the number of computations in the classic realm surpasses the quantum realm. The smallest system we have for the SP model with *s=2* has *n=6* and *N=64* showing that even for the smallest system the quantum algorithm has lower quantum computation.

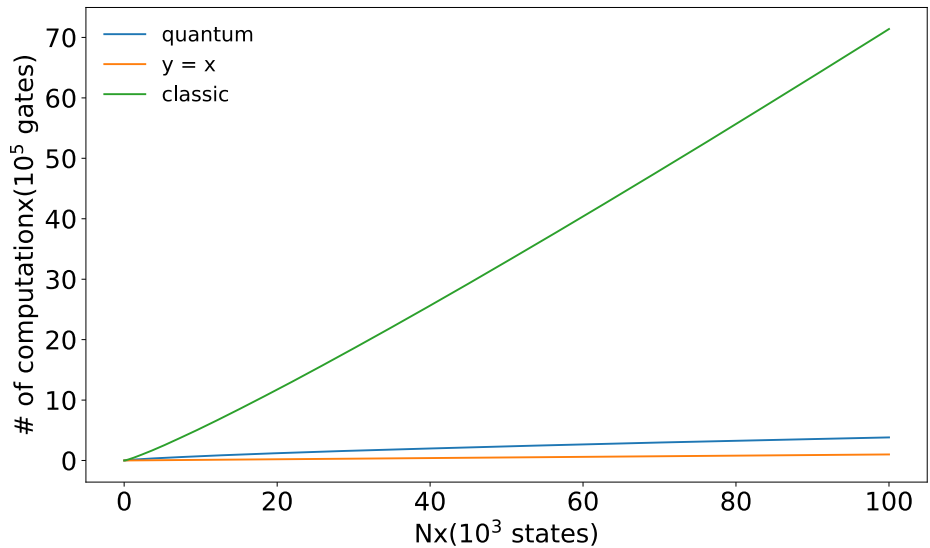

**FIG. S6**. Comparison between the number of gates for classic and quantum algorithms.

## V.III: MR model

For the MR model, the *#ofQ$_{init.}$* and the *#ofQ$_{diff.}$* are the same as the SP model. Even in the *#ofQ$_{orcl.}$* the introducing the energy, subtracting the $E_{tot}$ from $E_{th}$ and the negation steps are identical to the SP model. The only difference is that for the SP model the energy values in the energy table are simply added to find the $E_{tot}$, while in the MR model, these values are multiplied with the $d^{-1}$. Since the same procedure with the same computational cost does occurs in the classical version of the MR model, the general computational cost will still be much higher for the classical model, compared to the quantum MR model.

## VI. IMPLEMENTING ENERGIES IN THE CIRCUIT

In this part, we show parts of a circuit where some pair-wise energies are implemented using quantum gates (Fig. S7). Here, we have two interacting residues (Fig. S7-A), with the energy table described in Fig. S7-B. Since we have two designable sites, we require $n = 6$ qubits, where each residue uses 3 qubits as shown in Fig. S7-C. In the circuit shown in Fig. S7-C, parts of the energy table are implemented through *multi–control–not gates*, specifically *6-control-1-NOT* gates.

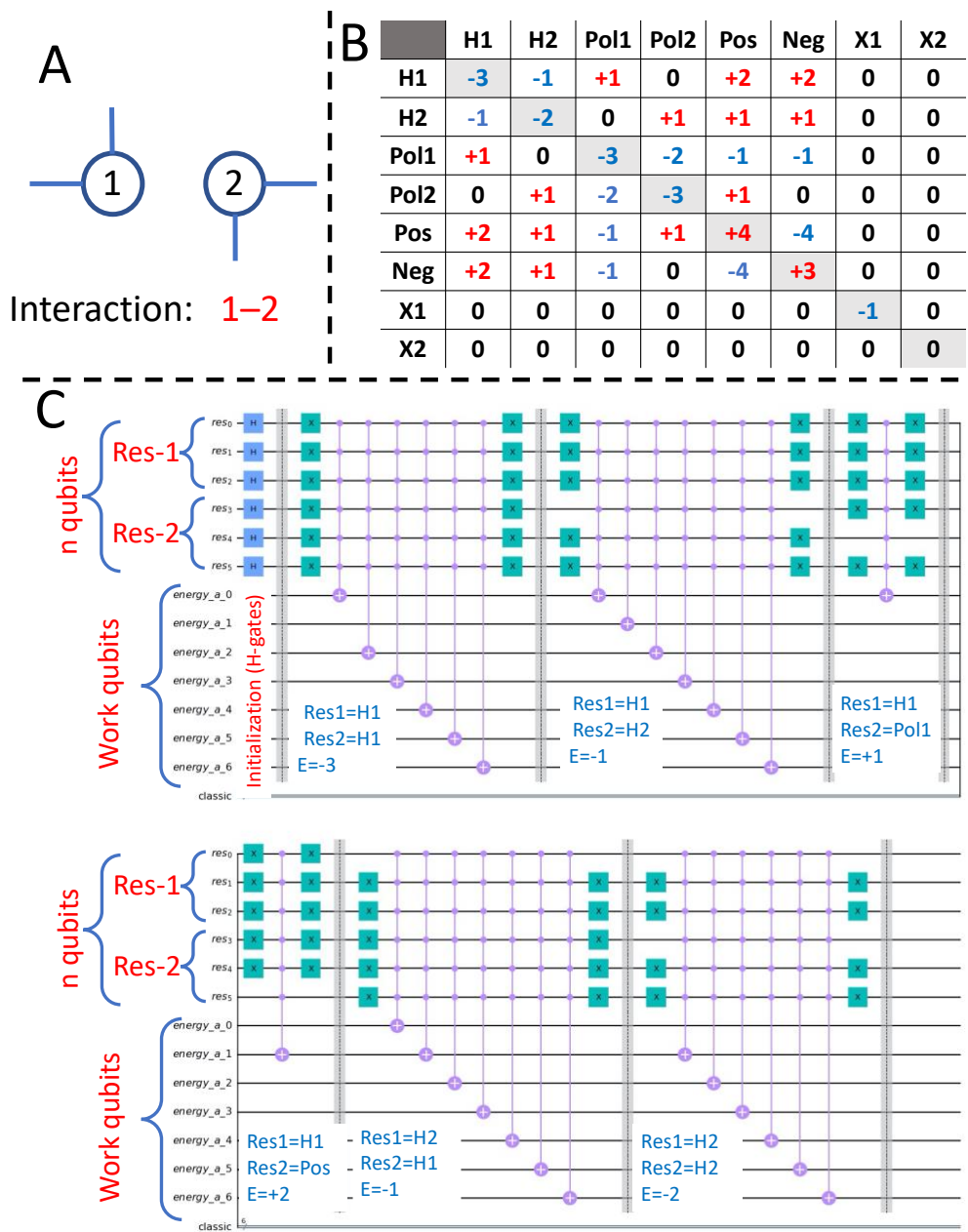

**FIG. S7**. A) representation of two interacting residues. B) energy table used in our study. C) Circuit representation of the initialization step and implementation of some of the pair-wise energies in the circuit, with *g=3*, *n=6* and *m=7*.

## VII. Computational Cost of Simulations and Circuit Depth

Despite obtaining the quantum advantage expected from Grover's algorithm (Fig. 6 in the main text), since we use quantum computer simulators to study these systems, running our circuits are restricted by the limitations of the conventional computers, i.e., the amount of RAM required to simulate a system and how long the simulation will take in real-time (CPU time). The changes in the amount of RAM and CPU time required to simulate different systems as a function of R show a linear increase for both the RAM usage and the CPU time for all data sets, which are provided in Supplementary section S-III. This linear increase is expected as by adding the number of iterations, the number of times the oracle and the diffuser are called in the algorithm are increased (Fig. 1 in the main text). Thus, since the simulations run on conventional computational resources, increasing the amount of computation linearly with R increases the cost of computation linearly.

Our results show that simpler circuits, i.e., *s=2* and *s=3* in the SP model and *s=2* in the MR model, could run with *R=$R_{max}$* on resources available on a regular laptop device. For these systems, the required RAM amount is ~8 GB, and the maximum CPU time is within an hour time range. However, for the rest of the circuits, simulating only the first few iterations could take days of simulation and more than 10 GB of RAM. For example, the *s=6,i=5,$E_{th}$=−19* system would need ~1.5 TB of RAM and ~71,392 hours (~8.1 years) of CPU time to simulate with *R = $R_{max}$=284* iterations. Moreover, for systems with the same number of designable sites and interactions, the computational resources required to simulate the circuits are almost identical for all *$E_{th}$* values, which indicates that the usage of computational resources is independent of the *$E_{th}$*.

Comparing the results between systems with similar configurations and results in the SP and MR model, i.e., the *s=2,$E_{th}$=−3* and the *s=2,$E_{th}$=95%$E_{min}$* circuits, show that the CPU time is more than 18 times longer and the RAM usage is twice more in the MR model circuits. These differences are expected as increasing the complexity of the oracle in the circuit increases the computational cost of simulations.

Figure S.8 provides insight into the depth of the quantum circuits and their effect on the CPU time and the RAM usage of simulations. Our results show that by increasing the circuit depth, the RAM usage and the CPU time for all simulations increase with similar but distinct power-law behaviours (Fig. S.8-a and b). This similarity confirms the consistency of circuits and gates among systems in each model. However, the different types of gate-based calculations performed in the oracle for each model (and the number of qubits required to execute them) lead to the distinct behaviour of the RAM usage and CPU time patterns in the SP and MR models. Moreover, the increase in the CPU time is faster for simulations in the MR model, while the RAM usage grows faster for systems in the SP model, which could be related to the way the simulator is programmed and is beyond the scope of this study.

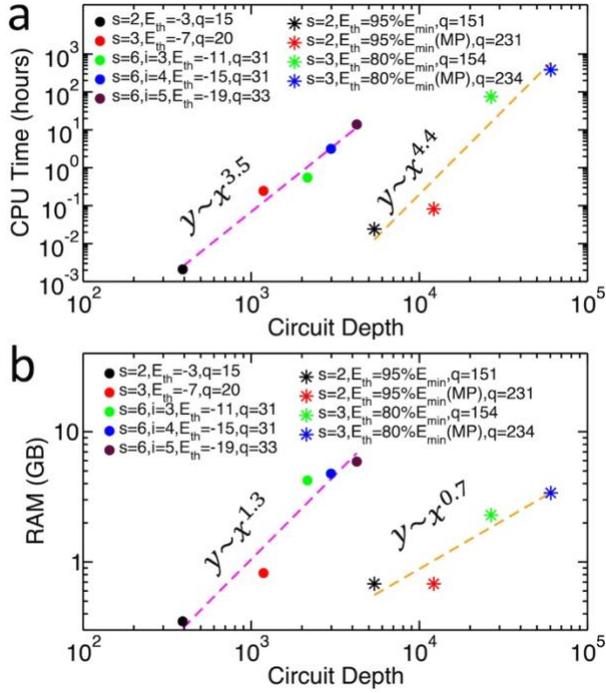

**FIG. S8:** Role of circuit depth in computational costs of simulating the circuits at R=1. Change in the: a) RAM usage; b) CPU time as a function of circuit depth. The dashed lines represent the fit to the data points. The q indicates the total number of qubits in the circuit.